\documentstyle[aps,preprint,epsfig,amsfonts,amsbsy,floats,times]{revtex}

\tightenlines
\draft

\begin{document}
\title{Dynamic surface critical behavior of isotropic Heisenberg ferromagnets}
\author{M.\ Krech$^1$, H.\ Karl$^2$, and H.\ W.\ Diehl$^2$}
\address{$^1$Fachbereich Physik, Bergische Universit\"at - GH  Wuppertal,
42097 Wuppertal, Federal Republic of Germany}

\address{$^2$Fachbereich Physik, Universit\"at - GH  Essen,
45117 Essen, Federal Republic of Germany}
\date{\today}
\maketitle

\begin{abstract}
The effects of free surfaces on the dynamic critical behavior
of isotropic Heisenberg ferromagnets are studied via
phenomenological scaling theory, field-theoretic renormalization group
tools, and high-precision computer simulations. An appropriate
semi-infinite extension of the stochastic model J is constructed, the
boundary terms of the associated dynamic field theory are identified,
its renormalization in $d\le 6$ dimensions is clarified, and the
boundary conditions it satisfies are given. Scaling laws are derived
which relate the critical indices of the dynamic and static infrared
singularities of surface quantities to familiar \emph{static} bulk and
surface exponents. Accurate computer-simulation data are presented for
the dynamic surface structure factor; these are in conformity with
the predicted scaling behavior and could be checked by appropriate
scattering experiments.
\end{abstract}

A key ingredient of modern theories of critical phenomena is
the arrangement of microscopically different systems
in universality classes \cite{MEF98}. First developed for
static equilibrium bulk critical behavior, this
classification scheme has subsequently been extended
both to \emph{dynamic} bulk critical behavior \cite{HH77} as well as to
\emph{static surface} critical behavior of semi-infinite systems at
\emph{bulk critical points} \cite{Bin83,Diehl}.

Since distinct dynamics may have the same equilibrium
distribution, each \emph{static} bulk universality class
generally splits up into \emph{several dynamic} ones,
represented by stochastic models called A, B, \ldots, J \cite{HH77}.
Likewise, to which \emph{static surface} universality class
a given system belongs is decided by its static bulk
universality class and additional relevant surface properties.
Hence each static surface universality class and each
dynamic bulk universality class generally splits up into
separate dynamic surface universality classes \cite{DJ92,DieWich}.

Unfortunately, detailed investigations of dynamic surface critical
behavior have remained scarce \cite{DJ92,DieWich,DD,FD,XiGo}
and largely focused on models with purely relaxational dynamics.
\emph{Isotropic Heisenberg ferromagnets} form an important class of
systems, characterized by the presence of \emph{nondissipative}
(mode-coupling) terms and a \emph{conserved} order parameter,
whose dynamic surface critical behavior has not yet been investigated.
To fill this gap, we shall use two different lines of approaches:
(i) phenomenological scaling and the field-theoretic renormalization
group (RG); (ii) computer-simulation studies of the dynamic surface
structure function.

Building on (i), we shall conclude that the critical indices
characterizing the dynamic surface critical  behavior of such systems
(which are $O(3)$ symmetric both  in the
bulk and at the surface) can be expressed  in terms of
\emph{known static} bulk and surface critical exponents.
Recently developed highly efficient spin dynamics algorithms
\cite{SD,DPLMK} have enabled us to corroborate these findings numerically.

In the simulations we utilized a
classical isotropic Heisenberg ferromagnet on a simple cubic lattice
whose sites $\boldsymbol{i}=(i_1,i_2,i_3)$, $i_1,i_2,i_3=0,\ldots,L-1$,
are occupied by spins $\boldsymbol{S}_{\boldsymbol{i}}=
(S^{\alpha}_{\boldsymbol{i}},\alpha{=}1,2,3)$ of
length $|\boldsymbol{S}_{\boldsymbol{i}}|=1$. Free boundary conditions
apply along the $i_3$ direction, and periodic ones along the others,
 so that the layers $i_3=0$ and $i_3=L-1$ are free surfaces.
The dynamics is defined through
\begin{equation}
\label{eqmot}
{d{\boldsymbol{S}}_{\boldsymbol{i}}\over dt}
 = {{\partial {{\mathcal H}}_{\text{lat}}} \over
\partial {\boldsymbol{S}}_{\boldsymbol{i}}}
\times {\boldsymbol{S}}_{\boldsymbol{i}}\;,
\end{equation}
with the Hamiltonian
\begin{equation}
\label{Hamil}
{\mathcal H}_{\text{lat}} = - J \sum_{\langle \boldsymbol{i},
\boldsymbol{j} \rangle \atop i_3\,\text{or}\,j_3\ne 0,L-1}
{\boldsymbol{S}}_{\boldsymbol{i}} {\cdot} {\boldsymbol{S}}_{\boldsymbol{j} }
- J_1 \sum_{\langle \boldsymbol{i},\boldsymbol{j} \rangle
\atop i_3=j_3= 0,L-1}
{\boldsymbol{S}}_{\boldsymbol{i}} {\cdot} {\boldsymbol{S}}_{\boldsymbol{j}}\;,
\end{equation}
whose nearest-neighbor bulk and surface bonds $J$ and $J_1$
(measured in units of temperature $T$) are ferromagnetic.

In the thermodynamic limit the model undergoes a continuous bulk phase
transition whose critical behavior is representative of the $O(3)$
universality class. Owing to the $O(3)$ symmetry, the surface of such
a $d{=}3$-dimensional system cannot spontaneously order for
$J_1/J<\infty$. Hence the surface transition that occurs at the bulk
critical point $T=T_{\text{c}}$ is the so-called \emph{ordinary} one
\cite{Diehl}. Its critical indices can be expressed in terms of two
independent bulk exponents, e.g., $\eta$ and $\nu$, and one surface
exponent, e.g., the correlation exponent $\eta^{\text{ord}}_{\|}$.

In our computer simulations, (\ref{eqmot}) is integrated
numerically for a given set of more than $ 700$ initial
spin configurations generated by a Monte-Carlo simulation
of the thermal equilibrium state  ${\propto}\, {\mathrm e}^{-{\mathcal
H}_{\text{lat}}}$ \cite{SD,DPLMK,MKslab}.
The spin-spin cumulant
\begin{eqnarray} \label{Sijtt}
C^{\alpha \beta}{\left(\boldsymbol{r}; z, z'; |t {-} t'| \right)}
\equiv {\big\langle S_{\boldsymbol{i}}^{\alpha}(t)\,
S_{\boldsymbol{i}'}^{\beta}(t')  \big\rangle}^{\text{cum}}
\end{eqnarray}
with $\boldsymbol{r}=( i_1{-}i_2,i'_1{-}i'_2)$, $z=i_3$, and $z'=i_3'$
is calculated, where $t$ and $t'$ are times to which the initial spin
configuration at $t=0$ has evolved according to (\ref{eqmot}). The
average $\langle{.}\rangle$ is taken over the set of initial
configurations. Before turning to the results, let us discuss
\emph{scaling}.

Consider a $d$-dimensional analog of the above model in
the thermodynamic limit $L\to\infty$.
According to scaling considerations \cite{HH69,Wag70} and RG work
\cite{MM75,BJW76}, the dynamic bulk critical exponent ${\mathfrak{z}}$
is given by
\begin{equation}\label{dynz}
{\mathfrak{z}}=\cases{(d+2-\eta)/2 &if $d \le d^*_{\text{J}}\equiv 6$.
\cr 4 & if $d > 6$.}
\end{equation}
Here  $d^*_{\text{J}} = 6$ is the upper critical dimension of model J,
and the mean-field (MF) value $\eta=0$ applies for
$d>d_{\text{st}}^*=4$, the static upper critical dimension.

Let $2 < d < 4$, so that hyperscaling holds.
A scaling ansatz for the cumulant (\ref{Sijtt}) at $T_{\text{c}}$
must comply with the scaling forms of (a) the bulk cumulant that
(\ref{Sijtt}) approaches as $z,z' \to \infty$ with $z{-}z'$
fixed, and (b) the static cumulant
$C{\left(\boldsymbol{r}; z, z'; 0\right)}$. Both requirements impose
strong restrictions. To see this, recall that the static critical behavior
is captured by the semi-infinite $3$-vector model
with Hamiltonian
\begin{equation}
\label{HGL}
{\mathcal H} = {\int_{\mathbb{R}^d_+}} \left[{1\over 2}\,
(\boldsymbol{\nabla} \boldsymbol{\phi})^2 + {\tau_0\over 2}\,\phi^2 +
{u_0\over 4!}\,|\boldsymbol{\phi}|^4\right]
+ {\int_{\mathcal B}}   {c_0\over 2}\, \phi^2 \;,
\end{equation}
defined on the half-space ${\mathbb{R}}^d_+
\equiv\{{(\boldsymbol{x}_{\|},z)}\in
{\mathbb{R}}^d\mid z\ge 0\}$ with boundary plane ${\mathcal B}$ at $z=0$.
In this model there is ordinary surface critical behavior at
bulk criticality if $c_0>c_{\text{sp}}$. At the corresponding
ordinary fixed point $c$, the renormalized analog of $c_0{-}c_{\text{sp}}$,
takes the value $c=\infty$. This implies that the order-parameter field
$\boldsymbol{\phi}$ asymptotically satisfies the Dirichlet boundary
condition $\left.\boldsymbol{\phi}\right|_{\mathcal B}=0$. The behavior
of $\boldsymbol{\phi}(\boldsymbol{x})$ as
$\boldsymbol{x}\equiv(\boldsymbol{x}_{\|},z)$
approaches the surface point
${\boldsymbol{x}}_{\mathcal B}\equiv(\boldsymbol{x}_{\|},0)$
follows from the boundary operator expansion (BOE)\cite{Diehl}
\begin{equation}\label{BOE}
\boldsymbol{\phi}({\boldsymbol{x}}) = D(z)\,
{\partial_{n}\boldsymbol{\phi}}(\boldsymbol{x}_{\mathcal B})+\ldots\;.
\end{equation}
The scaling dimensions of the operators $\boldsymbol{\phi}$
and ${\partial_{n}\boldsymbol{\phi}}$ are
$(d{-}2{+}\eta)/2$ and $(d{-}2{+}\eta^{\text{ord}}_{\|})/2$, respectively,
whence $D(z)=D_0\,z^{(\eta^{\text{ord}}_{\|}-\eta)/2}\;$.

From requirement (b) we draw the important conclusion that
$\boldsymbol{\phi}\mid_{\boldsymbol{x}\notin {\mathcal B}}$ and
the surface operator $\partial_{n}\boldsymbol{\phi}$
ought to retain their scaling dimensions in the dynamic case.
The reason is that for the locally scale invariant theory considered
here the scaling dimension of any local scaling operator ${\mathcal
O}(\boldsymbol{x},t)$ cannot differ from its bulk value except for
points $\boldsymbol{x}$ on the boundary \cite{Diehl}
or at an initial condition \cite{JSS89}. However, the initial
configurations $\boldsymbol{\phi}_{t{=}0}$ are drawn from the
equilibrium distribution and should therefore \emph{not} give rise
to different scaling dimensions.

Requirement (a) tells us that the dependence on $t$ must involve the
usual bulk scaling variables, such as $t r^{-\mathfrak{z}}$.
Additional relevant or marginal dynamic surface scaling fields can be
ruled out because (i) at the \emph{ordinary} transition the surface
does \emph{not} act as an independent source of critical behavior
\cite{Diehl} and (ii) the $O(3)$ invariance of the interactions
precludes a local violation of the order-parameter conservation at the
surface. Hence, sufficiently close to $T_{\text{c}}$, possible times
scales generated by the surface should be much smaller than the
characteristic bulk time scale. An appropriate scaling ansatz for
$C^{\alpha\beta}=\delta_{\alpha \beta} C$ at $T=T_{\text{c}}$ thus is
\begin{equation}\label{Czzpt}
C(\boldsymbol{r};z,z';t)=
r^{2-d-\eta}\,\Upsilon{\left({z/ r},{z'/ r};t{r}^{-{\mathfrak{z}}}\right)}\;.
\end{equation}
The BOE (\ref{BOE}) leads to
$\Upsilon\left({\mathsf{z}},{\mathsf{z}}';{\mathsf{t}}\right) \approx
\left({{\mathsf{z}}{\mathsf{z}}'} \right)^{(\eta_{\|}^{\text{ord}}-\eta)/2}\,
{\Upsilon_0}({\mathsf{t}})$ in the limit ${\mathsf{z}},{\mathsf{z}}'\to 0$.
By consistency we must have
\begin{eqnarray}\label{C11}
{\hat{C}_{11}}(\boldsymbol{p},\omega) &=&
p^{\eta_{\|}^{\text{ord}}-1-{\mathfrak{z}}}\, {\sigma}{\left(\omega
p^{-{\mathfrak{z}}}\right)}\;,\\
\label{C0}
{\hat{C}_{11}}(\boldsymbol{0},\omega) &=& \text{const}\;
\omega^{-{\left({\mathfrak{z}}+1-\eta_{\|}^{\text{ord}}\right)}/
{\mathfrak{z}}}\,,
\end{eqnarray}
where
${\hat C}_{11}(\boldsymbol{p},\omega)$,
the dynamic surface structure function,
is the Fourier transform of $C(\boldsymbol{r};0,0;t)$
with respect to $\boldsymbol{r}$ and $t$.

For $4<d<6$, the breakdown of hyperscaling (caused by the dangerous
character of the irrelevant scaling field $u\sim u_0$) must be taken
into account. While static exponents like $\eta$ and
$\eta_{\|}^{\text{ord}}$ take their MF values $0$ and $2$,
respectively, the value $(d{+}2)/2$ of $\mathfrak{z}$ agrees with the
MF value $3$ of the magnetic shift exponent  only for $d=4$.

To put the above findings on a firmer basis, let us
construct an appropriate semi-infinite extension of model J.
We may assume that the surface-induced modifications of both
the interactions and the dynamics are restricted
to the immediate vicinity of the
boundary ${\mathcal B}$. Hence, for points with $z>0$,
we use the stochastic bulk equation \cite{remark}
\begin{equation}\label{stocheqJ}
\dot{\boldsymbol{\phi}}(\boldsymbol{x},t)-\boldsymbol{\zeta}
=\lambda_0\,{\left(\triangle\,
\frac{\delta{\mathcal H}}{\delta\boldsymbol{\phi}}
+f_0\,\frac{\delta{\mathcal H}}{\delta\boldsymbol{\phi}}
\times{\boldsymbol{\phi}}\right)}\;,
\end{equation}
where $\boldsymbol{\zeta}$ is a Gaussian random force
with average $\langle\boldsymbol{\zeta}\rangle=0$ and
\begin{equation}\label{stocheq}
{\left\langle{\zeta^\alpha}(\boldsymbol{x},t)\,
{\zeta^\beta}(\boldsymbol{x}',t')
\right\rangle}=-\lambda_0\,{\delta^{\alpha\beta}}
\triangle\,\delta(\boldsymbol{x}{-}\boldsymbol{x}')\,\delta(t{-}t')\;.
\end{equation}
The derivative ${\delta{\mathcal H}\over{\delta\boldsymbol{\phi}}}$
involves a contribution $\delta(z) ({c_0}{-}\partial_z)\phi$
implied by the boundary term of $\delta{\mathcal H}$, which yields
one obvious surface contribution to (\ref{stocheqJ}).
But there may be others, corresponding to local changes of
the dynamics. As expounded in \cite{DJ92}, the requirements
of detailed balance, locality, order-parameter conservation,
absence of irrelevant and redundant operators, and here also
$O(3)$ symmetry, impose strong constraints, which are best dealt with
on the level of the equivalent path-integral formulation (see, e.g.,
\cite{BJW76}). To ensure detailed balance, the action
must have the form
\begin{equation}
\label{J}
{\mathcal J} = {\int_{-\infty}^{\infty}}\!dt\,
{\int_{\boldsymbol{x}}}
{\left\{\tilde{\boldsymbol{\phi}}\cdot\!
{\left[\dot{\boldsymbol{\phi}}
+ \boldsymbol{{\mathcal R}}{\cdot}{\left(
{\delta {\mathcal H} \over \delta \boldsymbol{\phi}}
-\tilde{\boldsymbol{\phi}}\right)}-
\frac{\delta\boldsymbol{\mathcal R}}{\delta\boldsymbol{\phi}}
\right]} \right\}},
\end{equation}
where $\int_{\boldsymbol{x}}$ comprises both volume and surface
integrals, $\tilde{\boldsymbol{\phi}}$ is the usual auxiliary field
needed in such a Lagrangian formulation, and a prepoint
discretization in time is used.

The right-hand side of (\ref{stocheqJ}) can be written as
$-{\boldsymbol{\mathcal R}}\cdot
\frac{\delta{\mathcal H}}{\delta\boldsymbol{\phi}}$ with
\begin{equation}\label{reactop}
{\mathcal R}^{\alpha\beta}=-\lambda_0\,{\left(\delta^{\alpha\beta}\triangle
+f_0\,{\epsilon^{\alpha\beta\gamma}}\,\phi^{\gamma}\right)}\;.
\end{equation}
To find out which surface terms must be included in ${\mathcal J}$
and the associated boundary conditions, we proceed as in
Refs.~\cite{DJ92,DieWich}. As reaction operator
in the action (\ref{J}) we use (\ref{reactop}), with the Laplacian
$\triangle$ replaced by
$-{\loarrow{\boldsymbol{\nabla}}}{\cdot}
{\roarrow{\boldsymbol{\nabla}}}$ (where $\loarrow{\boldsymbol{\nabla}}$
acts to the left). Contributions to
${{\mathcal R}}^{\alpha\beta}/\lambda_0$ of the form
${{\tilde{c}}_0^{\alpha\beta}}\delta(z)$ correspond to
\emph{nonconservative} dynamic surface terms \cite{DieWich}
and are ruled out by the presumed spin isotropy. As boundary
conditions (valid in an operator sense \cite{Diehl,DJ92}) we obtain
\begin{eqnarray}
{\left(\partial_n-c_0\right)}\boldsymbol{\phi}&=&0\,,\label{bcphi1}
\\\partial_n
{\delta{\mathcal H}\over\delta \boldsymbol{\phi}}=
\partial_n{\left(\tau_0+\frac{u_0}{6}\,
|\boldsymbol{\phi}|^2-\triangle\right)}{\boldsymbol{\phi}}
&=&0\,,\label{bcphi2}
\\\partial_n\tilde{\boldsymbol{\phi}}&=&0\,,\label{bcphit1}
\\{\left(\partial_n-c_0\right)}\tilde{\boldsymbol{\phi}}}
{\cdot} {\loarrow{\boldsymbol{{\mathcal R}}}
=\lambda_0\,{\left(c_0-\partial_n\right)}
{\left(\triangle\tilde{\boldsymbol{\phi}}-f_0\,
\tilde{\boldsymbol{\phi}}{\times}\boldsymbol{\phi}\right)}&=&0\,,
\label{bcphit2} 
\end{eqnarray}
where $\partial_n$ (=$\partial_z$) is a derivative along the inner normal.
The first, (\ref{bcphi1}), is known from statics.
The second, (\ref{bcphi2}), means that 
$\boldsymbol{n}{\cdot}\boldsymbol{j}^{\alpha}$, the normal component of 
the current $\boldsymbol{j}^{\alpha}=-\lambda_0\,(
\boldsymbol{\nabla}{\delta{\mathcal H}\over\delta\phi^\alpha}
+f_0\,\epsilon^{\alpha\beta\gamma}\phi^\beta\boldsymbol{\nabla}\phi^\gamma)$,
whose negative divergence gives the right-hand side of (\ref{stocheqJ}),
vanishes at ${\mathcal B}$. [The precession
term yields a contribution
$\propto \epsilon^{\alpha\beta\gamma}\phi^\beta\partial_n\phi^\gamma$
which is zero by (\ref{bcphi1})].
The remaining two
ensure self-adjointness of $\triangle$
and consistency with the
fluctuation-dissipation theorem (FDT)
\begin{equation}\label{FDT}
-\theta(t) \langle{\dot{\phi}^\alpha}(\boldsymbol{x},t)
{\phi^{\beta}}{(\boldsymbol{x}',0)} \rangle
={\langle{\phi^\alpha}{(\boldsymbol{x},t)}(\tilde{\boldsymbol{\phi}}}{\cdot}
{\loarrow{\boldsymbol{\mathcal{R}}})^\beta(\boldsymbol{x}',0)\rangle}.
\end{equation}

To study the ordinary transition, it is convenient to set
$c_0{=}\infty$. Then (\ref{bcphi1}) and (\ref{bcphit2}) simplify to
Dirichlet boundary conditions for $\boldsymbol{\phi}$ and
${\tilde{\boldsymbol{\phi}}}{\cdot}{\loarrow{\boldsymbol{{\mathcal R}}}}$. In
applying the RG, we consider two distinct renormalization schemes: a
massless one, RS$_1$, based on the $\epsilon$ expansion about
$d^*_{\text{J}}$, and a massive one, RS$_2$, in fixed dimension $d\le 4$.
Common to both is the form of the reparametrizations
$\boldsymbol{\phi}=Z_{\phi}^{1/2}{\boldsymbol{\phi}_{\text{ren}}}$,
$\tilde{\boldsymbol{\phi}}=
Z_{\phi}^{-1/2}\tilde{\boldsymbol{\phi}}_{\text{ren}}$,
$\partial_n\boldsymbol{\phi}=[Z_{\phi}Z_{1,\infty}]^{1/2}
[\partial_n{\boldsymbol{\phi}]_{\text{ren}}}$,
$\lambda_0=\mu^{-4}{Z_\lambda}\lambda$, and
$f_0=\mu^{3-{d/ 2}}Z_{\phi}^{1/2}Z_\lambda^{-1}f$,
where $\mu$ is either an arbitrary momentum scale (RS$_1$) or
the renormalized mass (RS$_2$)%
. That the $Z$-factors of $\tilde{\boldsymbol{\phi}}$ and
$\boldsymbol{\phi}$ can be chosen reciprocal to each other
follows from the fact that the bulk vertex function
$\Gamma^{\text{(b)}}_{\tilde{\phi}\phi}$ has no primitive uv
divergence $\propto \delta'(t{-}\tilde{t})$. The form of the
$Z$-factor of $f$ is a consequence of (\ref{FDT}) \cite{BJW76}; it
implies that the beta function
$\beta_f\equiv{\left.\mu\partial_\mu\right|_0}f$ becomes
$\beta_f=\frac{1}{2}(d{-}6{-}\eta_\phi{+}2\eta_\lambda)f$,
with $\eta_{\phi,\lambda}\equiv
{\left.\mu\partial_\mu\right|_0}\ln Z_{\phi,\lambda}$.
From the resulting exact relation
$\eta_\lambda^*=\frac{1}{2}(6{-}d{+}\eta_\phi^*)$
among the values
$\eta_\phi^*=\eta$ and $\eta_\lambda^*=4-\mathfrak{z}$
at the infrared-stable fixed point, expression (\ref{dynz})
for $\mathfrak{z}$ follows.

Consider first RS$_1$ for $4<d\le 6$. For $u_0=0$,
the theory is renormalizable and
$Z_\phi{=}Z_{1,\infty}{=} 1$ to all orders in $f$. Only one
bulk renormalization factor $Z_\lambda$ remains
whose expansion to order $f^2$ may be gleaned from
\cite{BJW76}, which also tells us that single insertions of the composite operator $\tilde{\boldsymbol{\phi}}{\times}\boldsymbol{\phi}$
appearing in
${\tilde{\boldsymbol{\phi}}}{\cdot}{\loarrow{\boldsymbol{{\mathcal R}}}}$
additionally require a subtraction
$\propto \Delta \tilde{\boldsymbol{\phi}}$.
Together with this latter counterterm, the above reparametrizations
suffice to cure the uv singularities of the correlation
and response functions \cite{Kar00}, including those involving the surface
operators $\partial_n\boldsymbol{\phi}$ and
$\tilde{\boldsymbol{\phi}}|_{z{=}0}$ and single insertions of 
$\tilde{\boldsymbol{\phi}}{\times}\boldsymbol{\phi}$.
The proof \cite{proof} utilizes power counting,
the boundary conditions (\ref{bcphi1})--(\ref{bcphit2}), and the FDT
(\ref{FDT}) \cite{expl}. The BOE (\ref{BOE}) holds with $D(z)\sim z$.

If $u_0\ne 0$ and $d< 4$, perturbative massless RG schemes are
plagued by infrared singularities \cite{Sym73}.
The usual escape via an $\epsilon$ expansion here
is not possible because of the different upper critical dimensions
$6$ and $4$ associated with $f_0$ and $u_0$.
We therefore use RS$_2$.  Writing $u_0=\mu^{d-4}Z_uu$,
we fix all static  bulk and surface counterterms (mass shift
$\tau_0{-}\mu^2$, $Z_{\phi}$, $Z_u$, $Z_{1,\infty}$)
as in \cite[Eqs.\ (3.3a-d), (710a,b)]{DS94}. The remaining dynamic (bulk)
counterterms ($Z_\lambda$, subtraction for
$\tilde{\boldsymbol{\phi}}{\times}\boldsymbol{\phi}$)
can be fixed by appropriate
(massive) normalization conditions for the dynamic bulk theory.
From the resulting RG equations of the dynamic theory
and well-established RG results for the static theory
the BOE (\ref{BOE}) and the scaling forms (\ref{Czzpt})--(\ref{C0})
follow.

\begin{figure}[htb]
\centerline{\epsfig{figure=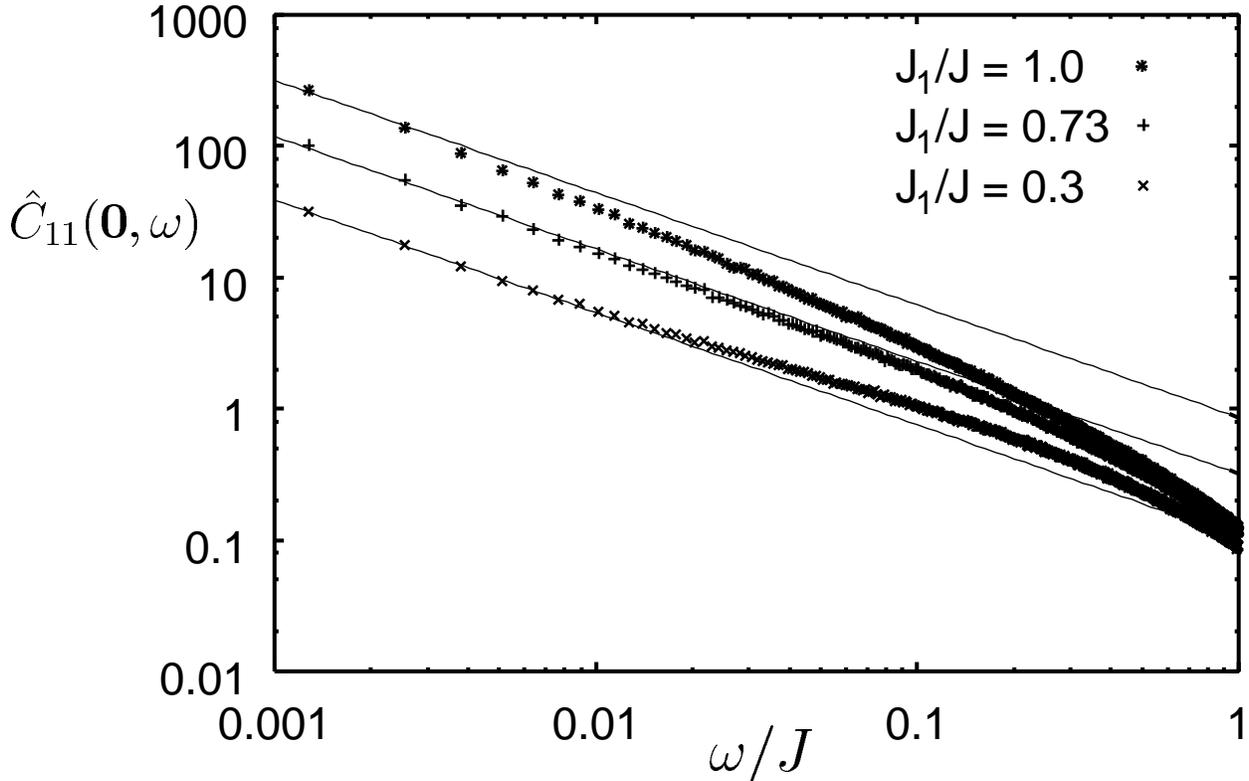,width=\columnwidth}}
\caption{Structure function
$\hat{C}_{11}(\boldsymbol{0},\omega)$
for $J_1/J$ = 0.3 $(\times)$, 0.73 $(+)$, and 1.0 $(*)$.
Error bars (one standard deviation) are smaller than the symbol sizes.
The solid lines indicate the theoretically expected power law
(\protect\ref{C0}) for $\omega \to 0$.
\label{S10wfig}}
\end{figure}

\begin{figure}[htb]
\centerline{\epsfig{figure=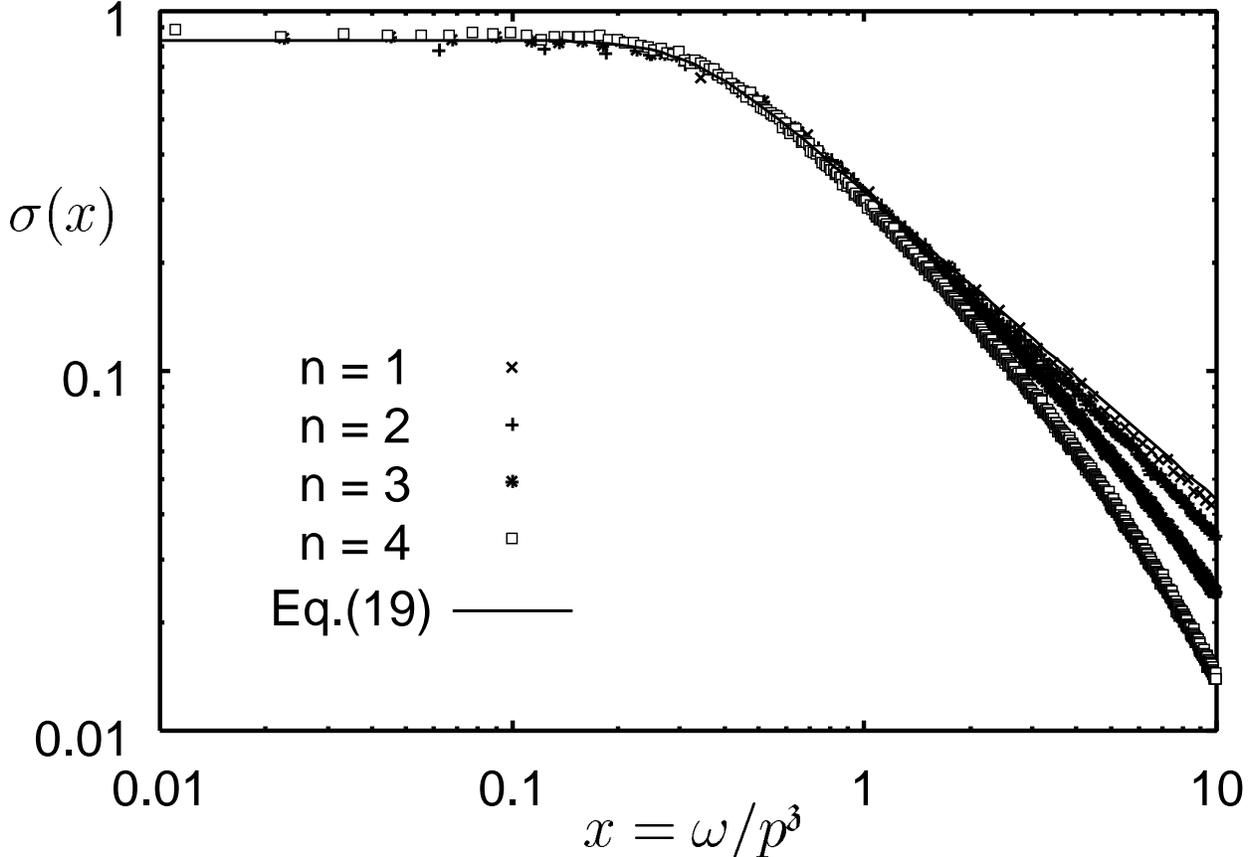,width=\columnwidth}}
\caption{Scaling function $\sigma(x)$ according to
(\protect\ref{C11}). Data obtained for $J_1/J = 0.73$ and
$\boldsymbol{p}=({n\pi\over 30},0,0)$, with $n= 1,\ldots,4$, are
shown. Error bars (one standard deviation) are smaller than the symbol
sizes. The solid line displays a fit to (\protect\ref{sigma}). The
data for $x \geq 1$ are outside the scaling regime.
\label{S1pw}}
\end{figure}

Our simulation results are displayed in Figs.~\ref{S10wfig}
and \ref{S1pw}. They were obtained for $T = T_{\text{c}}$,
$L = 60$, and a total integration time of about $5000/J$.
The algorithm was parallelized and implemented on the ALiCE
cluster at the BUGH Wuppertal, where the MPI (message passing
interface) library was used for communication between the processes.
Finite-size effects turned out to be negligible
for this choice of parameters.

In Fig.~\ref{S10wfig} the structure function
$\hat{C}_{11}(\boldsymbol{0},\omega)$ is shown for different values of
$J_1/J$ in comparison with (\ref{C0}). For the exponent
$({\mathfrak{z}} {+} 1 {-} \eta_{\|}^{\text{ord}})/{\mathfrak{z}}$, we
used the value $0.856\,{\pm}\, 0.005$ that follows from the estimate
${\mathfrak{z}}(d{=}3)=2.482\,{\pm}\,0.002$, obtained by substitution
of the value \cite{GZJ98} $\eta(d{=}3)=0.036\,{\pm}\, 0.004$ into
(\ref{dynz}), and the current estimate \cite{MKslab} 
$\eta^{\text{ord}}_{\|}(d{=}3)=1.358 \,{\pm}\, 0.012$.

Depending on whether $J_1/J$ is small $(J_1/J{=}0.3, \times)$ or 
larger $({J_1/J}{=}1, *)$, the approach to the asymptotic power law
(\protect\ref{C0}) is from above or below. In the latter case,
the asymptotic regime is {\em not} reached within the frequency range
shown in Fig.~\ref{S10wfig}. The best agreement with (\ref{C0})
over the largest frequency range is obtained for the intermediate
value $J_1/J = 0.73\, (+)$.

Fig.~\ref{S1pw} shows a scaling plot of
$\hat{C}_{11}({\boldsymbol{p}},\omega)$ for
$\boldsymbol{p} = ({n\pi\over 30},0,0)$.
The scaling regime in $x$ shrinks as
the mode index
$n$ is increased from $1 (\times)$ to $4 (\Box)$.
The shape of the scaling function in (\ref{C11}) , for $x < 1$, is
captured surprisingly well  by
\begin{equation}
\label{sigma}
\sigma(x) = \sigma_0
\left[ 1 + (x/x_0)^4
\right]^{(\eta^{\text{ord}}_{\|}-{\mathfrak{z}}-1)/4{ \mathfrak{z}}}\;,
\end{equation}
whose form is inspired by the known zero-loop result \cite{DJ92,DieWich}. 
The exponent at the  square bracket is chosen so as to reproduce
(\ref{C0}) in the limit $x \to \infty$ ($p \to 0$ at fixed
$\omega\ne 0$). The amplitude $\sigma_0$ and the crossover parameter
$x_0$ are used as fit parameters. The fit shown in Fig.~\ref{S1pw} has been
obtained from the data for $n = 3$ and $x < 1$.

The agreement between the data displayed in Figs.~\ref{S10wfig}
and \ref{S1pw} and the scaling laws (\ref{C11}) and (\ref{C0}) is
quite satisfactory, albeit small deviations are seen to
remain on closer inspection. Just as in \cite{MKslab},
intermediate values of $J_1/J$ ($= 0.73$) 
appear to be empirically optimal in that they yield
the largest asymptotic scaling regime.
Possible sources of deviations are (i) insufficient momentum
resolution, (ii) insufficient frequency resolution, and (iii) corrections
to scaling. Momentum resolution is intimately linked to the system
size $L$, which despite formidable progress in simulation techniques,
still is a serious limiting factor. Frequency resolution is limited by the
total integration time. From
inspection of $C_{11}(\boldsymbol{p},t)$ (not shown) we conclude that the
integration time is sufficiently long. The frequency resolution $\delta
\omega / J \simeq 1.2 {\times}10^{-3}$ available here rivals
that of neutron scattering experiments \cite{DPLMK}.
Momentum resolution is more modest.

Corrections to scaling
may have familiar static roots or be of genuine dynamic origin.
Examples of the former category are corrections $\sim (u{-}u^*)$
governed by the static Wegner exponent $\omega_u\equiv\beta_u'(u^*)
\simeq 0.8$, for $d=3$ \cite{GZJ98}, and corrections due to the finiteness
of $c_0$ (`nonzero extrapolation length') \cite{Diehl}.
An example of the second category are Wegner-type corrections
$\sim (f{-}f^*)$, governed by $\omega_f\equiv \beta_f'(u^*,f^*)$,
the dynamic analog of $\omega_u$, whose $\epsilon$ expansion about
$d{=}6$ reads $\omega_f=\epsilon + O(\epsilon^2)$ \cite{BJW76}.
While we cannot rule out that corrections of the latter category are large,
we are not aware of compelling reasons to expect this.
Thus the mentioned corrections of static origin may well be
the most important ones.

In summary, we have presented a detailed study of the surface critical
behavior of isotropic Heisenberg ferromagnets based on
phenomenological scaling and field-theoretical RG methods
and corroborated our findings through high-precision simulations.
Our results depicted in Figs.~\ref{S10wfig} and \ref{S1pw}
may serve as guidelines for careful experimental tests.
For these, one must choose ferromagnets for which
dipolar forces (ignored here) can be trusted to be unimportant.

We thank K.\ Wiese for helpful discussions. M.K.\ is
indebted to the {\em Institut f\"ur angewandte Informatik} at the
BUGH Wuppertal for providing access to the parallel cluster ALiCE.
Partial support by DFG (for M.K.\ via the Heisenberg program under
grant \# Kr 1322/2-1, for H.K.\ and H.W.D.\ via the Leibniz program
and SFB 237) is gratefully acknowledged.


\begin{thebibliography}{99}
\bibitem{remark}  Since model J differs from the lattice model 
 (\ref{eqmot}) in that the energy is not a conserved density,
it may seem more appropriate to consider model J',
which involves coupling to a conserved density.
However, this coupling is expected to be irrelevant.
Clear indications are: (i) models B and D
(= models J and J' with $f_0=0$)
have the same value
$4{-}\eta$ of ${\mathfrak{z}}$; (ii) the Gaussian value $(d{+}2)/2$
of ${\mathfrak{z}}$ associated with the precession term is
smaller than  its analog, $4$, for models  B and D, whenever $2<d<6$.
\end{thebibliography}
\end{document}